\documentclass[12pt]{elsarticle}
\usepackage{amsmath,amssymb}
\usepackage{amsfonts}
\usepackage{bm}

\begin{document}

\begin{frontmatter}

\title{Hamiltonian formulation of the effective kinetic theory for superfluid Fermi liquids}
\author{Manuel Valle}
\ead{manuel.valle@ehu.es}
\address{Departamento de F\'\i sica Te\'orica, 
 Universidad del Pa\'\i s Vasco, 
 Apartado 644, E-48080 Bilbao, Spain }

\begin{abstract}
We present in a local form the time dependent effective description of a superfluid Fermi liquid which includes Landau damping effects at $T\neq 0$. 
This is achieved by the introduction of an additional variable, the quasiparticle distribution function, which
obeys a simple kinetic equation. 
The transport equation is coupled with first order equations for the Goldstone mode and the particle density.
We prove that a main feature of this formulation is its Hamiltonian structure relative to 
a certain Poisson bracket. We construct the Hamiltonian to quadratic order.   
\end{abstract}


\end{frontmatter}

\section{Introduction}

The effective description at finite temperature of the broken symmetry phase  
of a Fermi gas in the BCS-BEC crossover is complicated by the 
effect of Landau damping, which results in a highly nonlocal 
time-dependent Ginzburg-Landau theory~\cite{Abrahams}. 
Apart from the fact that the derivation of such an effective Lagrangian at $T\neq 0$ 
for either the complex order parameter or 
only the Goldstone mode is rather tricky even at 
quadratic level~\cite{Stoof,Aitchison1,Aitchison2,Alamoudi,Benfatto}, 
this nonlocal formulation makes it difficult to simulate numerically 
the real time evolution of non-equilibrium processes, 
such as oscillations of trapped Fermi gases. 

Indeed, a similar situation arises in dealing with collective effects and  dynamical screening 
of Abelian and non-Abelian plasmas at high temperature or high density. 
To leading order in the coupling constant, consistency requires the inclusion of  a set 
of one-loop diagrams termed ``hard thermal loops'', which are derived from 
a nonlocal effective Lagrangian~\cite{Braaten}. 
Fortunately, the equations of motion can be written in a local form by introducing 
auxiliary fields~\cite{Blaizot0,Nair0}. In the Abelian case, the auxiliary field 
obeys a linearized Vlasov equation corresponding to the collisionless regime in the plasma. 
Remarkably, it turns out that the resulting equations form a Hamiltonian system 
with  a noncanonical bracket  structure~\cite{Nair,Iancu}. 

In this paper, we turn to the question of the derivation of the low-energy dynamics of the 
phase of the order parameter in the superfluid phase,  in connection with the above analogy.  
We present  a simple derivation of the linearized equations of motion  in a local form. 
Moreover, 
we prove that this local formulation is Hamiltonian, and $H$  and the Poisson structure 
are completely identified.

\section{The equations of motion}
\label{motion}

We begin by introducing the action for the system. In terms 
of the Nambu spinor $\Psi^\dagger = (\psi_\uparrow^\dagger, \psi_\downarrow)$
the action for the two-component  balanced  Fermi system is  written as 
\begin{eqnarray}
S &=&   \int d\bm{x}\int dt  \,\Psi^\dagger(X) \left[ i \partial_t + \tau^3 \frac{1}{2 m} \nabla^2  + \tau^3 \mu
-\tau^3 V_\mathrm{ext}(X) \right.  \nonumber \\ 
&&+ \left. \tau^+  \Delta(X) +\tau^-  \Delta^\ast(X)  \right]  \Psi(X)  
-\int d\bm{x}\int dt  \,\frac{1}{g_\Lambda}  \Delta^\ast \Delta , 
\end{eqnarray} 
where $\tau^{\pm} = \frac{1}{2} \left(\tau^1 \pm i \tau^2 \right)$, 
and the $\tau^j$'s are Pauli matrices.
Here $\mu$ is the chemical potencial,  $g_\Lambda$ is the bare coupling parameter, 
and we have included an arbitrary trapping  potential $V_\mathrm{ext}(X)$. 
The complex field
$\Delta(X)$  performs the 
Hubbard-Stratonovich decoupling of the quartic interaction between fermions in the 
BCS channel.

In order to derive an effective theory   
when the symmetry $U(1)$ is broken to $Z_2$,  
it is convenient to express the above Lagrangian in terms of  
a Goldstone field $\theta(X)$ and
``heavy''  fields, $\widetilde{\Psi}(X)$ and $\widetilde{\Delta}(X)$  to be integrated out
\begin{eqnarray}
\Psi(X) &=&  e^{i \tau^3 \theta(X)} \widetilde{\Psi}(X), \\
\Delta(X) &=& e^{2 i \theta(X)} \widetilde{\Delta}(X).    
\end{eqnarray}
Let us choose the non-zero expectation value of the scalar field as real, 
$\langle \Delta \rangle= \Delta_0$. 
The condition that $\widetilde{\Delta}(X)$ does not 
contain the Goldstone mode\footnote{For a detailed discussion on how to construct effective 
Lagrangians in the case of spontaneously  broken symmetries, see Weinberg~\cite{broken}.}  
turns out to be $\mathrm{Im} \left( \widetilde{\Delta}(X)^\ast \Delta_0 \right) = 0$, 
so the heavy field $\widetilde{\Delta}(X)$ needs to be real with arbitrary sign. 
In  terms of the covariant derivative
\begin{equation}
D_\mu \widetilde{\Psi}(X) = \partial_\mu \widetilde{\Psi}(X) +
 i \partial_\mu \theta(X) \tau^3 \widetilde{\Psi}(X) , 
\end{equation}
the Lagrangian  becomes 
\begin{eqnarray}
\mathcal{L} &=& \widetilde{\Psi} i D_t \widetilde{\Psi} - 
 \frac{1}{2m}\left( D_i \widetilde{\Psi}\right)^\dagger \tau^3 D_i \widetilde{\Psi} 
 + \widetilde{\Psi}^\dagger \tau^1 \widetilde{\Psi} \, \widetilde{\Delta} - 
 \frac{1}{g_\Lambda} \widetilde{\Delta}^2 \nonumber \\
 && +
 \widetilde{\Psi}^\dagger \tau^3 \widetilde{\Psi} \,\left(\mu- V_\mathrm{ext} \right) ,
\end{eqnarray}
and the Noether current of the $U(1)$ symmetry, which now is realized as 
$\delta\widetilde{\Psi} = \delta\widetilde{\Delta}=0$ and 
$\delta\theta = \epsilon$,  
 is given by  
\begin{eqnarray}
J^0 &=& n = \widetilde{\Psi}^\dagger \tau^3  \widetilde{\Psi}, \\  
\label{eq:JJ}
J^k &=&  \frac{1}{2 m i} \left( \widetilde{\Psi}^\dagger \partial_k\widetilde{\Psi}  - 
                                            \partial_k \widetilde{\Psi}^\dagger \widetilde{\Psi} \right)  +
                                            \widetilde{\Psi}^\dagger \tau^3  \widetilde{\Psi}\, \frac{\partial_k \theta}{m} . 
\end{eqnarray}

To effectively integrate out the fermionic degrees of freedom we note that
the total Hamitonian for these 
may be viewed as $H = H_0 + H_\mathrm{ext}$, where 
\begin{equation}
H_0 = \int d\bm{x} \widetilde{\Psi}^\dagger \left( 
-\tau^3 \frac{\nabla^2}{2 m}  - \tau^3 \mu - \tau^1 \Delta_0 \right) \widetilde{\Psi} , 
\end{equation}
and 
\begin{eqnarray}
 H_\mathrm{ext} &=& \int d\bm{x} \left(  
  \widetilde{\Psi}^\dagger \tau^3 \widetilde{\Psi} (\partial_t \theta + V_\mathrm{ex})
- \widetilde{\Psi}^\dagger \tau^1 \widetilde{\Psi} \, {\sigma}\right. \nonumber \\ 
&&+ \left. \frac{1}{2 m i} \left( \widetilde{\Psi}^\dagger \bm{\nabla} \widetilde{\Psi}  - 
                                            \bm{\nabla} \widetilde{\Psi}^\dagger \widetilde{\Psi} \right) 
                                            \cdot \bm{\nabla}\theta  + 
                                            \widetilde{\Psi}^\dagger \tau^3  \widetilde{\Psi}\, 
                                            \frac{(\bm{\nabla} \theta)^2}{2m} \right) .                   
\end{eqnarray}
Therefore $H_\mathrm{ext}$ couples the system to an  
applied perturbation given by 
 $\sigma(X) \equiv \widetilde{\Delta}(X) - \Delta_0$,  the gradients of $\theta$,  and 
 $V_\mathrm{ext}$. 
 Following the procedure reviewed in Ref.~\cite{Blaizot}  
 in the framework of the high temperature regime of QCD, 
 the idea is to compute the  induced changes in the expectation values  
 of $n$, $\bm{J}$ and the pairing field 
 $\widetilde{\Psi}^\dagger \tau^1 \widetilde{\Psi}$ by the external perturbations 
 $\{ \partial_\mu \theta, \sigma, V_\mathrm{ext} \}$. 
 By ignoring nonlinear corrections, we need the retarded response functions 
 $\chi_{AB}(X) = -i \langle[A(X),B(0)] \rangle \theta(t)$  whose Fourier transforms 
are denoted by $\chi_{AB}(Q)$.  
In the linear response approximation the induced changes take the form~\cite{Manes} 
 \begin{eqnarray}
 \label{eq:dn}
\delta \langle n(Q)\rangle &=& -\chi_{n n}(Q) \upsilon(Q) + \chi_ {n J}^{ \;\; k}(Q) i q^k \theta(Q)  -
 \chi_{n 1}(Q) \sigma(Q), \\ 
\delta \langle J^k(Q)\rangle &=& -\chi_{J n}^{k}(Q)\upsilon(Q) + 
  \chi_ {J J} ^{k \; l}(Q) i q^l \theta(Q) -
 \chi_{J 1}^{k} (Q) \sigma(Q) , \\
\delta \langle \Psi^\dagger \tau^1 \Psi \rangle 
 &=& -\chi_{1 n}(Q) \upsilon(Q) + \chi_ {1 J}^{\; \; k}(Q) i q^k \theta(Q)  -
 \chi_{1 1}(Q) \sigma(Q), 
\end{eqnarray} 
where 
\begin{equation}
\upsilon(Q) = i \omega \theta(Q) - V_\mathrm{ext}(Q) . 
\end{equation}  
From a functional point of view, if $\Gamma^{(2)}[\theta, \sigma]$ is the 
quadratic approximation for the effective action that one obtains after integrating out the fermions, 
and $S[\sigma] =  -g_\Lambda^{-1} \int  \sigma^2$ is the bare action, 
then  the integration of the heavy 
field $\sigma$ is simply gaussian.
This produces an effective action for the Goldstone mode of the form  
$\Gamma^{(2)}[\theta, \bar{\sigma}] + S[\bar{\sigma}] - \frac{1}{2}
\ln \mathrm{det}\left( -2 g_\Lambda^{-1} - \chi_{11} \right) $, where $\bar{\sigma}$ is the solution to the 
saddle point condition $\delta(S[\sigma]  + \Gamma^{(2)}[\theta, \sigma])/\delta\sigma =0$. 
The last term, related to the determinant, does not depend on $\theta$, but  gives rise to 
important corrections 
to the thermodynamic properties evaluated in the  
mean field approximation~\cite{Diener,Randeria}. 
However, for the purpose of deriving an effective description only in terms of the phase $\theta$, 
such a term  
will be ignored. 
Thus, in this approximation, the 
$\sigma$ field may be effectively integrated out by simply adjusting 
its value to the solution of the gap equation
\begin{equation}
 \label{eq:gap1}
\frac{2 \sigma}{g_\Lambda} = \delta \langle \widetilde{\Psi}^\dagger \tau^1 \widetilde{\Psi} \rangle  .
\end{equation}
Exploiting the fact that $n$ and $\theta$ are canonically conjugated variables, and imposing
conservation of the Noether current  
 once $\sigma$ has been eliminated,  
 we may derive a set of equations  for the 
 time derivatives of 
 $\delta\langle n\rangle$ and $\theta$. 
 These equations encode the effective dynamics.  
   
 The explicit expressions of the response functions  are easily computed at small frequency and momentum. 
 By keeping the $O(\lambda^0)$ terms after the scalings $\omega \to \lambda\omega$, 
 $\bm{q} \to \lambda \bm{q}$,  the general form of $\chi_{AB}(Q)$ when $T\neq 0$  is
  the sum a  regular piece independent of $Q$,  
  and a contribution due to Landau damping   which is non-analytical at $ Q=0$:  
  \begin{equation}
{\chi}_{AB}(Q)=\bar{\chi}_{AB}+
2 \int \frac{d^3 k}{(2\pi)^3}  F_{AB}(\bm{k}) 
\frac{\bm{q}\cdot \bm{\nabla}_k E_k \,  n_F'(E_k)}{\omega + i \eta - \bm{q}\cdot \bm{\nabla}_k E_k  } . 
\end{equation}
The non-vanishing regular parts are given by 
\begin{eqnarray}
\label{eq:nn}
\bar{\chi}_{n n} &=&-\int \frac{d^3 k}{(2 \pi)^3} \frac{\Delta_0^2}{E_k^3}
\tanh \frac{\beta E_k}{2}, \\ 
\label{eq:n1}
\bar{\chi}_{n 1} &=&\bar{\chi}_{1 n}= -\int \frac{d^3 k}{(2 \pi)^3} \frac{\Delta_0 \xi_k }{E_k^3}
 \tanh \frac{\beta E_k}{2}, \\ 
\bar{\chi}_{1 1}^{(\Lambda)} &=&-\int^\Lambda \frac{d^3 k}{(2 \pi)^3} 
 \frac{\xi_k^2}{E_k^3}\tanh \frac{\beta E_k}{2},  \\
\label{eq:jj1}
\bar{\chi}_ {J J} ^{k \; l} & = & -\frac{\hat{q}^k \hat{q}^l}{m}  
\int \frac{d^3 k}{(2 \pi)^3} \left(\frac{\xi_k}{E_k} \tanh \frac{\beta E_k}{2}- 1\right) = \frac{\langle n\rangle}{m}  \hat{q}^k \hat{q}^l  ,   
\end{eqnarray} 
with the standard notation,  $E_k = \sqrt{\xi_k^2 + \Delta_0^2}$,  $\xi_k =k^2/2m - \mu$. 
The form of  $\bar{\chi}_ {J J} ^{k \; l} $ is entirely due to the last term in Eq.~(\ref{eq:JJ}). 
On the other hand, the factors in the integrand of the Landau damping contributions are 
given by  
\begin{eqnarray}
F_{n n}(k) &=& -\frac{\xi_k^2}{E_k^2},  \\ 
F_{n 1}(k)&=&  F_{1 n}(k)= \frac{\Delta_0 \xi_k}{E_k^2}, \\ 
F_{1 1}(k) &=& -\frac{\Delta_0^2}{E_k^2},  \\ 
F_{J J}^{k\; l}(\bm{k}) &=& -\frac{k^k k^l}{m^2}, \\
F_{J n}^{l}(\bm{k}) &=& F_{n J}^{\; \;l}(\bm{k})= -\frac{\xi_k k^l}{m E_k}, \\ 
F_{J 1}^{l}(\bm{k}) &=& F_{1 J}^{\; \;l}(\bm{k})= \frac{\Delta_0 k^l}{m E_k}.
\end{eqnarray}
These last results determine the nonlocal part of  the departure 
from the equilibrium values of the quantities of interest.  
For instance, the nonlocal contribution to $\delta \langle n(X)\rangle$ 
in space-time 
is obtained by combining the above terms according to Eq.~(\ref{eq:dn}) 
\begin{equation}
\label{eq:deltan}
\delta \langle n(Q)\rangle^\mathrm{nonlocal} = 
-2 \int \frac{d^3 k}{(2\pi)^3} \frac{\xi_k}{E_k}   
\frac{\bm{q}\cdot \bm{\nabla}_k E_k \,  n_F'(E_k)}{\omega + 
i \eta - \bm{q}\cdot \bm{\nabla}_k E_k} \delta E(Q, \bm{k}),
\end{equation}
where 
\begin{eqnarray}
\delta E(Q, \bm{k}) &=& \frac{\partial E_k}{\partial \mu} \upsilon(Q) + 
 \frac{\partial E_k}{\partial \Delta_0} \sigma(Q) + 
i \bm{k} \cdot \bm{q}\frac{\theta(Q)}{m} \nonumber \\
&=& -\frac{\xi_k}{E_k}  \upsilon(Q) + \frac{\Delta_0}{E_k}  \sigma(Q) + 
i \bm{k} \cdot \bm{q}\frac{\theta(Q)}{m} . 
\end{eqnarray}
 
In order to present the equations in a local form, it is natural to introduce a new variable 
$w(Q; \bm{k})$, defined as 
\begin{equation}
\delta \langle n(Q)\rangle^\mathrm{nonlocal} \equiv 
2 \int \frac{d^3 k}{(2 \pi)^3} \frac{\xi_k}{E_k} w(Q;\bm{k}), 
\end{equation}
which measures the departure of the distribution function from equilibrium when $T\neq 0$. 
From this definition and Eq.~(\ref{eq:deltan}) it follows that 
\begin{equation}
\left( \omega - \bm{q}\cdot \bm{\nabla}_k E_k  \right) w(Q;\bm{k}) = 
-n_F'(E_k) \bm{\nabla}_k E_k \cdot  \bm{q} \, \delta E(Q, \bm{k}), 
\end{equation}
and the equation of motion for distribution function $w(X;\bm{k})$ takes the form of a 
transport equation without collision term 
resembling a linealized Vlasov equation
\begin{equation}
\label{eq:vlasov}
(\partial_t  + \bm{\nabla}_k E_{k} \cdot \bm{\nabla}_x )w(X, \bm{k}) =  
 n_F'(E_k) \bm{\nabla}_k E_k \cdot \bm{\nabla}_x \delta E(X, \bm{k})  , 
\end{equation}
where $\delta E$ is the induced change in the energy of the quasiparticle due to the 
applied perturbation
\begin{equation}
\delta E(X,\bm{k}) = -\frac{\xi_k}{E_k} \upsilon(X) + \frac{\Delta_0}{E_k} \sigma(X) + 
\bm{k} \cdot \frac{\bm{\nabla} \theta(X)}{m} .
\end{equation} 
Thus,  by combining the expressions for $F_{AB}(\bm{k})$ with the definition of $w$, 
one finds the  total  changes  in a local form
\begin{eqnarray}
\label{eq:dn0}
\delta\langle n(X)\rangle &=& -\bar{\chi}_{nn} \upsilon(X) - 
 \bar{\chi}_{n1} \sigma(X) + 
 2 \int \frac{d^3 k}{(2 \pi)^3} \frac{\xi_k}{E_k} w(X;\bm{k}), \\ 
\label{eq:deltaJ}
\delta\langle J^i(X) \rangle &=& \frac{\langle n \rangle}{m} \nabla_i \theta  + 
 2 \int \frac{d^3 k}{(2 \pi)^3} \frac{k^i}{m} w(X;\bm{k}), \\ 
 \label{eq:d1}
\delta\langle \Psi^\dagger \tau^1 \Psi(X)\rangle &=& 
 -\bar{\chi}_{1 1}^{(\Lambda)} \sigma(X) - 
 \bar{\chi}_{n1} \upsilon(X) - 
 2 \int \frac{d^3 k}{(2 \pi)^3} \frac{\Delta_0}{E_k} w(X;\bm{k}). 
\end{eqnarray}

The transport equation~(\ref{eq:vlasov}) was derived 
long time ago by Betbeder-Matibet and Nozi\`eres~\cite{Nozieres}, 
and more recently by Urban and Schuck~\cite{Urban}. 
In these approaches the starting point is the Heisenberg equation of motion 
for a matrix distribution function, whose diagonalization leads to the above kinetic equation. 
As Leggett~\cite{Legget1,Legget2} and Betbeder-Matibet and Nozi\`eres~\cite{Nozieres} have shown, 
it is possible to consider Fermi-liquid effects 
by adding to $\delta E(X,\bm{k})$ an extra term, $\sum_{\bm{k}'}f_{\bm{k} \bm{k}'} w(X,\bm{k}')$, 
where $f_{\bm{k} \bm{k}'}$ describes interactions of two elementary excitations. 

To eliminate the $\sigma$ field we use the gap equation~(\ref{eq:gap1}). 
The relation between the $s$-wave scattering length $a_s$ 
and the bare coupling constant $g_{\Lambda}$
\begin{equation}
 -\frac{m}{4 \pi a_s} 
 = \frac{1}{g_\Lambda} - \int^\Lambda \frac{d^3 k}{(2 \pi)^3} \frac{1}{2 \epsilon_k},
\end{equation}
together with the gap equation for $\Delta_0$ at $T\neq 0$, 
\begin{equation}
-\frac{m}{2\pi a_s} = \int \frac{d^3 k}{(2 \pi)^3} \left(\frac{1}{E_k}\tanh \frac{\beta E_k}{2} - 
\frac{1}{\epsilon_k}\right),
\end{equation}
produce  $\bar{\chi}_{1 1}^{(\Lambda)} =
 -2 g_\Lambda^{-1} -\bar{\chi}_{n n}$.
Combining this result with Eqs.~(\ref{eq:gap1}) and~(\ref{eq:d1}) 
we obtain\footnote{The term proportional to $\upsilon$ was omitted in~\cite{Nozieres}.}
\begin{equation}
\label{eq:sigma}
\sigma(X) = \frac{\bar{\chi}_{n 1}}{\bar{\chi}_{n n}} \upsilon(X) + 
\frac{2}{\bar{\chi}_{n n}} \int \frac{d^3 k}{(2 \pi)^3} \frac{\Delta}{E_k} w(X;\bm{k}).
\end{equation} 

Finally, the conservation of the particle number
\begin{equation}
 \partial_t \delta\langle n(X)\rangle + 
 \bm{\nabla} \cdot \delta\langle\bm{J}(X) \rangle = 0,
\end{equation}
yields the equation of motion satisfied by  $\upsilon$, thus completing the set of dynamical equations 
for the variables $\{w(X;\bm{k}), \theta(X), \upsilon(X)\}$. 
This is found to be 
\begin{eqnarray}
\label{eq:conserv}
&&-\bar{\chi}_{n n} \left(1 + \frac{\bar{\chi}_{n 1}^2}{\bar{\chi}_{n n}^2} \right) \frac{\partial \upsilon}{\partial t}
-2\int \frac{d^3 k}{(2 \pi)^3} \left(\frac{\xi_k}{E_k}  - \frac{\bar{\chi}_{n 1}}{\bar{\chi}_{n n}} \frac{\Delta_0}{E_k}\right)
\bm{\nabla}_k E_k \cdot \bm{\nabla}_x w(X,\bm{k})  \nonumber \\
&&\quad +2\int \frac{d^3 k}{(2 \pi)^3} \left(\frac{\xi_k}{E_k}  - \frac{\bar{\chi}_{n 1}}{\bar{\chi}_{n n}} \frac{\Delta_0}{E_k}\right)
n_F'(E_k)\bm{\nabla}_k E_k \cdot \bm{\nabla}_x \left( \bm{k} \cdot \frac{\bm{\nabla} \theta(X)}{m} \right) \nonumber  \\ 
&&\quad+ \frac{\langle n \rangle}{m} \nabla^2 \theta  + 
2\int \frac{d^3 k}{(2 \pi)^3} \frac{\bm{k}}{m}\cdot \bm{\nabla}_x w(X,\bm{k}) = 0, 
\end{eqnarray}
where we have used  Eqs.~(\ref{eq:vlasov}) and (\ref{eq:sigma}),  as well as the vanishing of the angular integration  
$\int d \Omega_{\bm{k}} \bm{\nabla}_k E_k f(k)$ for any 
isotropic function $f(k)$ such as $\xi_k/E_k$ and $\Delta_0/E_k$. 
Thus, the dynamical  equations~(\ref{eq:vlasov}), (\ref{eq:conserv}), together with 
\begin{equation}
\frac{\partial \theta}{\partial t} =  -\upsilon(X) - V_\mathrm{ext}(X), 
\end{equation}
and the equation~(\ref{eq:sigma}) for $\sigma(X)$ 
form a  closed set of local equations for  the effective low energy theory of the superfluid Fermi liquid at 
the one-loop level, when Fermi liquid effects are ignored.  

\section{Hamiltonian formulation}

As we have pointed out before, 
a similar situation to the one posed by the Landau damping terms in the present context  
also appears 
when one attempts  to derive a local time-dependent effective Lagrangian  
for the soft degrees of freedom of a gauge theory at high temperature or density.  
In that case, at the expense of introducing a new kind of  degrees of freedom, 
one can reformulate the equations of motion as a system of local equations. 
The new variables represent  the  fluctuations of the charged (or coloured) particle distributions, and  
satisfy  kinetic equations with external  and  
induced fields due to these fluctuations~\cite{Blaizot0}. 
It turns out that the complete set of dynamical equations  are Hamiltonian 
with respect to certain Poisson brackets~\cite{Nair, Iancu}. 
It is natural to ask whether a similar formulation can be given in this case, and if so,  
what the Hamiltonian structure describing  the low energy theory of the Fermi superfluid would be. 

To address this question, it is convenient to replace the variable $\upsilon(X)$ by the more natural
$n_1(X) \equiv \delta\langle n(X) \rangle$, and to consider $\{w(X;\bm{k}), \theta(X), n_1(X)\}$ as 
the set of dynamical variables. This choice exploits the fact that 
the particle density and the Goldstone mode are 
canonically conjugated, 
as follows from the role of the particle number operator as 
the generator of the 
$U(1)$ symmetry,  and the inhomogeneous 
transformation law for $\theta$,  $\delta\theta \propto i [N,\theta]$.   
Therefore the Poisson bracket of these variables may be written as
\begin{equation}
\label{eq:canon}
\left\{n_1(t, \bm{x}), \theta(t, \bm{y}) \right\}= \delta(\bm{x}-\bm{y}) . 
\end{equation}

The remainder Poisson structure may be guessed by noting that 
a kinetic equation of Vlasov type usually takes the Hamiltonian form
\[
\partial_t f(t, \bm{x},\bm{k})  = \left\{f, H \right\} .  
\] 
Here  the relevant bracket is the Poisson-Vlasov bracket~\cite{Marsden}
\begin{equation}
\label{eq:poi}
\left\{F[f], G[f] \right\} = \int d^3 x\, d^3 k  \, f 
\left\{\frac{\delta F}{\delta f} , \frac{\delta G}{\delta f}\right\}_{\bm{x} \bm{k}} ,
\end{equation}
where $\left\{, \right\}_{\bm{x} \bm{k}} $ is the canonical bracket on 
the single particle phase space $\mathbb{R}^6$ spanned by $(\bm{x}, \bm{k})$, 
\begin{equation}
\left\{f , g\right\}_{\bm{x} \bm{k}} \equiv  \bm{\nabla}_x f \cdot  \bm{\nabla}_k g - 
    \bm{\nabla}_x g \cdot  \bm{\nabla}_k f . 
\end{equation}
The functions $f, g$ are distribution functions on  $\mathbb{R}^6$.  
Since  we are interested in 
the linearized equations around equilibrium, where 
$f = n_F(E_k) + w(\bm{x},\bm{k})$,  
the Poisson structure will be  chosen as  
\begin{eqnarray}
\label{eq:PB3}
    \left\{F, G \right\}& = & P \int d^3 x\int d^3 k \, n_F(E_k)
     \left\{\frac{\delta F}{\delta w}, \frac{\delta G}{\delta w} \right\}_{\bm{x} \bm{k}}   \nonumber \\ 
    & &+ \int d^3 x 
     \left( 
    \frac{\delta F}{\delta n_1} \frac{\delta G}{\delta \theta} - 
    \frac{\delta G}{\delta n_1} \frac{\delta F}{\delta \theta} \right) . 
\end{eqnarray}
The first term  corresponds to a frozen Lie-Poisson bracket~\cite{Marsden}  which is 
obtained from~(\ref{eq:poi}) by linearization; 
the value of the factor $P$ will be determined shortly. 
The Jacobi identity for the frozen bracket follows from the general properties of 
Lie-Poisson brackets~\cite{Marsden}.
The second line corresponds to the  canonical Poisson structure of Eq.~(\ref{eq:canon}). 

It remains to see that the equations of motion we have found before  
are Hamilton's equations
\begin{eqnarray}
\label{eq:dtw}
\partial_t w &=& \left\{w, H \right\}  = 
P\,  n_F'(E_k) \bm{\nabla}_k E_k \cdot \bm{\nabla}_x \left(
\frac{\delta H}{\delta w(X,\bm{k})} \right) ,    \\ 
\partial_t \theta &=& \left\{\theta, H \right\}=  -\frac{\delta H}{\delta n_1(X)} ,   \\  
\label{eq:dtn}
\partial_t n_1 &=& \left\{n_1, H \right\} =  \frac{\delta H}{\delta \theta(X)} , 
\end{eqnarray}
for some quadratic functional $H[w, \theta, n_1]$. 
With the aid of the relations  
\begin{eqnarray}
\upsilon(X) &=&\frac{\bar{\chi}_{nn}}{\bar{\chi}_{nn}^2 + \bar{\chi}_{n1}^2}  
 \left[ -n_1(X) \right.  \nonumber \\
 &&  + \left.
                      2 \int \frac{d^3 k}{(2\pi)^3} \left(\frac{\xi_k}{E_k} -\frac{\bar{\chi}_{n1}}{\bar{\chi}_{nn}}  \frac{\Delta_0}{E_k}\right) w(X, \bm{k}) \right] , \\ 
 \sigma(X) &=&\frac{\bar{\chi}_{nn}}{\bar{\chi}_{nn}^2 + \bar{\chi}_{n1}^2}  
 \left[ -\frac{\bar{\chi}_{n1}}{\bar{\chi}_{nn}} n_1(X) \right. \nonumber \\  
  && +  \left.                    2 \int \frac{d^3 k}{(2\pi)^3} \left(\frac{\Delta_0}{E_k} +\frac{\bar{\chi}_{n1}}{\bar{\chi}_{nn}}  \frac{\xi_k}{E_k}\right) w(X, \bm{k}) \right] ,                      
\end{eqnarray}
we may express the change in the energy of the quasiparticle in terms of the new variables
\begin{eqnarray}
\delta E(X,\bm{k}) &= &\frac{\bar{\chi}_{nn}}{\bar{\chi}_{nn}^2 + \bar{\chi}_{n1}^2}  
 \left(\frac{\xi_k}{E_k} -\frac{\bar{\chi}_{n1}}{\bar{\chi}_{nn}}  \frac{\Delta_0}{E_k}\right) 
  n_1(X) \nonumber \\ 
 && + \bm{k} \cdot \frac{\bm{\nabla} \theta(X)}{m} +\delta E^{(1)}(X,\bm{k})  , 
\end{eqnarray}
where $\delta E^{(1)}$ denotes the contribution linear in $w$
\begin{eqnarray}
 \delta E^{(1)}(X,\bm{k}) &=&  \frac{2\bar{\chi}_{nn}}{\bar{\chi}_{nn}^2 + \bar{\chi}_{n1}^2} 
 \left[-\frac{\xi_k}{E_k}  \int \frac{d^3 k'}{(2\pi)^3} \left( \frac{\xi_{k'}}{E_{k'} } -  
 \frac{\bar{\chi}_{n1}}{\bar{\chi}_{nn}}\frac{\Delta_0}{E_{k'} }\right) w(X,\bm{k}')  \right. \nonumber \\ 
 &&+ \left.  \frac{\Delta_0}{E_k}  \int \frac{d^3 k'}{(2\pi)^3} \left( \frac{\Delta_0}{E_{k'} } +  
 \frac{\bar{\chi}_{n1}}{\bar{\chi}_{nn}}\frac{\xi_{k'}}{E_{k'} }\right) w(X,\bm{k}') \right] .
\end{eqnarray}  
By comparing Eq.~(\ref{eq:dtn}) for $\partial_t n_1$ and Eq.~(\ref{eq:deltaJ}), 
it follows that the Hamiltonian 
must contain exactly the term 
\[
\int  d^3 x \int \frac{d^3 k}{(2\pi)^3}\, 2 \bm{k}\cdot \frac{\bm{\nabla}\theta(X)}{m} 
 w(X,\bm{k})  . 
\]
On the other hand, its derivative with respect to $w$ when inserted into the RHS of Eq.~(\ref{eq:dtw})  
produces 
\[
P \frac{2}{(2\pi)^3} n_F'(E_k) \bm{\nabla}_k E_k \cdot \bm{\nabla}_x \left(
\bm{k}\cdot \frac{\bm{\nabla}\theta(X)}{m} \right), 
\]
so,  in order to match the corresponding  piece of $\partial_t w$ in Eq.~(\ref{eq:vlasov}), $P$ must assume  the value  
$P = (2\pi)^3/2$. 
Now, since 
\begin{equation}
\frac{\delta}{\delta w(X,\bm{k})} 
\int  d^3 y \, d^3 q \,  
  \delta E^{(1)}(t ,\bm{y},\bm{q})w(t, \bm{y},\bm{q}) = 2\, \delta E^{(1)}(X,\bm{k}),
\end{equation}
one can check immediately  
by functional derivation that the Hamiltonian
\begin{eqnarray}
\label{eq:hamil}
H[w,\theta,n_1]&=& \int  d^3 x \left(
 -\frac{\bar{\chi}_{nn}}{\bar{\chi}_{nn}^2 + \bar{\chi}_{n1}^2}
\frac{n_1(X)^2}{2} \right.  \nonumber \\
&&+ \left. n_1(X) V_{\mathrm{ext}}(X) + 
\frac{\langle n \rangle}{2m} (\bm{\nabla}\theta(X))^2  \right)  \nonumber \\ 
 &&+ \int  d^3 x \int \frac{d^3 k}{(2\pi)^3}  2 \bm{k}\cdot \frac{\bm{\nabla}\theta(X)}{m} 
 w(X,\bm{k})    \nonumber \\ 
 &&+ \int  d^3 x \int \frac{d^3 k}{(2\pi)^3} \frac{2 \bar{\chi}_{nn}}{\bar{\chi}_{nn}^2 + \bar{\chi}_{n1}^2}
 \left(\frac{\xi_k}{E_k} - \frac{\bar{\chi}_{n1}}{\bar{\chi}_{nn}}\frac{\Delta_0}{E_k} \right)
 n_1(X) w(X,\bm{k}) \nonumber \\ 
 &&+\int  d^3 x \int \frac{d^3 k}{(2\pi)^3} \left( -\frac{w(X,\bm{k})^2}{n_F'(E_k)}   
  +\delta E^{(1)}(X,\bm{k})w(X,\bm{k}) \right)  , 
\end{eqnarray} 
and the bracket~(\ref{eq:PB3}) 
yield the equation of  motion  for $\{w,\theta, n_1\}$. 
Although the Poisson structures are decoupled,  the effective Hamiltonian 
contains terms mixing  all variables.
As $n_F'(E_k)$ is a monotonic function the regularity of the integrand is guaranteed.  

When $\Delta_0 \to 0$  the coefficients $\bar{\chi}_{nn}$ and  $\bar{\chi}_{n1}$ vanish, and 
the most singular terms of the Hamiltonian are grouped as
\[
 -\frac{\bar{\chi}_{nn}}{\bar{\chi}_{nn}^2 + \bar{\chi}_{n1}^2}\frac{1}{2}
  \int  d^3 x \left( n_1 - 2 \int \frac{d^3 k}{(2\pi)^3}  \mathrm{sign}(\xi_k) w \right)^2 . 
\] 
To keep the energy finite, the change in the particle density for vanishing $\bar{\chi}$'s 
is restricted to
$n_1 = 2 \int_k  \mathrm{sign}(\xi_k) w$, 
according with Eq.~(\ref{eq:dn0}). 
Thus the equation of motion for $\theta$ yields $\partial_t \theta=0$, and
this variable in no longer time dependent.  
Now the Hamiltonian becomes
\begin{eqnarray}
H[w,\theta(\bm{x})]&=& -\int  d^3 x \int \frac{d^3 k}{(2\pi)^3}  \frac{w(X,\bm{k})^2}{n_F'(\xi_k)} 
 \nonumber \\ 
 &&+ 
2 \int  d^3 x V_{\mathrm{ext}}(X) \int \frac{d^3 k}{(2\pi)^3}  \mathrm{sign}(\xi_k) w(X,\bm{k}) \nonumber \\
&& + 
\int  d^3 x
\frac{\langle n \rangle}{2m} (\bm{\nabla}\theta(\bm{x}))^2 \nonumber \\    
&&+ 2\int  d^3 x \int \frac{d^3 k}{(2\pi)^3}  \bm{k}\cdot \frac{\bm{\nabla}\theta(\bm{x})}{m} 
 w(X,\bm{k})    , 
\end{eqnarray} 
where we have used $n_F'(|\xi_k|)=n_F'(\xi_k)$. 
The equation of motion for $\partial_t n_1$ obtained by variation of $\theta$ 
is not an independent equation as it corresponds exactly to  
the integration of 
$2\int _k  \mathrm{sign}(\xi_k) \partial_t w$. 
If  the initial condition is chosen  $\theta(t=0,\bm{x}) = \mathrm{constant}$, 
the $\bm{\nabla}\theta(\bm{x})$-terms in the Hamiltonian may be ignored, and
one recovers the Hamiltonian of a non-interacting Fermi gas in an external potential
\footnote{The term proportional to $w^2$ in Eq.~(\ref{eq:hamil}) is similar to one 
used in~\cite{Kruskal} to express the variations of the energy of a plasma in equilibrium.}.
By contrast, in the limit of zero temperature when $n_F'(E_k) \to 0$, the distribution function $w$ must vanish  in order 
to keep  the energy finite. Therefore one recovers a description 
in terms of $n_1$ and $\theta$ alone. 

\section{Conclusion}

To conclude, we have provided the Poisson structure and an effective Hamiltonian 
for the low energy description of  a superfluid Fermi gas in the collisionless regime. 
The basic dynamical variables  are the quasiparticle distribution function, the Goldstone mode, and the 
particle number density. 
Apart from the limitations inherent to the linear approximation to the equations of motion we have made, 
the most serious limitation of this approach is that it neglects Fermi liquid effects.   
A more accurate treatment would require the addition  of the contribution  
$1/2 \int d\bm{k}\, d\bm{k}' \,  f_{\bm{k} \bm{k}'} w(X,\bm{k})w(X,\bm{k}')$ 
in terms of the appropriate Landau parameters to the above effective Hamiltonian.

\section*{Acknowledgments}

I  thank I\~nigo Egusquiza and Juan L. Ma\~nes for helpful discussions. 
This work is supported by  the Spanish 
Ministry of Science and Technology 
under Grant  FPA2009-10612, 
the Spanish Consolider-Ingenio 2010 Programme CPAN (CSD2007-00042) and 
the Basque Government under Grant No. IT559-10.

\bibliographystyle{model1a-num-names}
\bibliography{<your-bib-database>}

\end{document}